\documentclass{article}

\usepackage{html}
\usepackage{makeidx}
\usepackage{RR}

\makeindex

\RRauthor{Thierry Despeyroux}

\RRdate{D\'ecembre 2006}

\RRetitle{Developing efficient parsers in Prolog:\\ the CLF manual (v1.0)}

\RRtitle{Programmer des analyseurs syntaxiques efficaces en Prolog: le manuel CLF (v1.0)}

\titlehead{CLF manual V1.0}

\RRresume{Ce document d\'ecrits deux outils utiles pour rapidement
concevoir et d\'evelopper des langages informatiques. Le premier utilise
Flex pour constuire des analyseur lexicaux, le second est une
extension des DCG de Prolog pour construire des analyseurs
syntaxiques. \`A l'origine con\c{c}u comme un nouveau composant du syst\`eme
Centaur, ces outils sont maintenant disponibles de fa\c{c}on
ind\'ependante. Les analyseurs produits peuvent \^etre int\'egr\'es dans des
syst\`emes \'ecrits en Prolog ou dans d'autres langages. Ceci est la
verion initiale du document, des versions mises \`a jour seront
disponible en ligne si n\'ecessaire}

\RRabstract{This document describes
a couple of tools that help to quickly design and develop computer
(formalized) languages. The first one use Flex to perform lexical
analysis and the second is an extention of Prolog DCGs to perfom
syntactical analysis. Initially designed as a new component for the
Centaur system, these tools are now available independently and can be
used to construct efficient Prolog parsers that can be integrated in
Prolog or heterogeneous systems. This is the initial version of the
CLF documentation. Updated version will be available online when necessary.}

\RRkeyword{lexical analysis, syntactical analysis, Lex, Prolog, DCG, Definite Clause Grammar}

\RRmotcle{analyse lexicale, analyse syntaxique, Lex, Prolog, DCG, Definite Clause Grammar}

\RRprojet{AxIS}

\RRtheme{\THCog}

\URRocq

\begin{document}

\newcommand{\pointer}[1]
{\htmladdimg{../pics/pointer.gif}
\hyperref{#1}{{\small $\triangleright$ {\it #1} }}{ {\small{\it p.}\pageref{#1}}}{#1}}

\newcommand{\pointerb}[2]
{\htmladdimg{../pics/pointer.gif}
\hyperref{#1, }{{\small $\triangleright$ {\it #1} }}{ {\small{\it p.}\pageref{#1}}}{#1}
\hyperref{#2}{{\small - {\it #2} }}{ {\small{\it p.}\pageref{#2}}}{#2}}

\newcommand{\pointerc}[3]
{\htmladdimg{../pics/pointer.gif}
\hyperref{#1, }{{\small $\triangleright$ {\it #1} }}{ {\small{\it p.}\pageref{#1}}}{#1}
\hyperref{#2}{{\small - {\it #2} }}{ {\small{\it p.}\pageref{#2}}}{#2}
\hyperref{#3}{{\small - {\it #3} }}{ {\small{\it p.}\pageref{#3}}}{#3}}

\makeRT

\tableofcontents


\section{Introduction}


The initial goal of the Computer Language Factory (CLF) was to enhance
the trilogy of formalisms Metal/PPML/Typol used in Centaur
\cite{BCDI88} to prototype the syntax and the semantics of computer
languages.  The Centaur system was a syntactic editor, written in
Lisp, and was able to call external modules as parsers or semantic
tools that was specified using specific formalisms.  However, with the
premature end of the development of the Centaur sytem, this goal was
not completely achieved.

The current version of CLF has been rebuild to permit a fast an easy
developpement of efficient parsers in Prolog, combining Flex and an
extended version of Prolog
DCGs\cite{ClocksinMellish,83204,learnprolognow}. It has been used in
the
AxIS\footnote{http://www.inria.fr/recherche/equipes/axis.en.html}
team to produce a parser for XML and some derived languages
\cite{Despeyroux:WWW2004}. These parsers has been used intensively to
parse a great number of Xhtml and XML files. As an exemple, parsing a
24 millions charaters (614000 lines) OWL ontology takes only 13
seconds on a 2 Mhz Pentium laptop with 1MB of memory, while loading of
the same triples takes 50 seconds with Protege.

One can imagine and use a computer formalism without any concrete
syntax. This can be the case for example for an intermediate code
formalism that does not need to be visualized. In this case only an
abstract syntax can be given. However most computer languages have a
concrete syntax. In the case of a translator we can want to define
only an unparser for the target language, but most of the time one
will need also a parser and thus a scanner. The parser is thus a very
important interface between the human and the computer representation
of programs or any other formalized documents.

As it was the case for Metal \cite{KLMM83} in the Mentor
\cite{DHKLL84} and Centaur \cite{BCDI88} systems, the CLF does not use
its own formalism to formalize a scanner, but uses an interface to
scanners generated with Flex.

Unless the traditional use of Lex and Yacc, we consider the process of
analysing some text as a translation from a list of tokens to a target
language. With this approach the lexical analyser and the scanner are
not linked together, so they can be developped and debbuged separatly.
On multi processors or multi cores systems, the lexical analyser and
the parser can be dispatched, improving performances for the user.

A great difference between the Centaur implementation and the current
one is that we do not use any abstract syntax declaration. So there is
no static analysis of the source code and the constructed term is a
completely free Prolog term.


\section{Rational}

\htmladdimg{../pics/RBline.gif}

As a preamble, we think that it is easier and safer to specify than to
directly program. This is in particular true when one want to modify
an existing program: a specification is in general much simpler as it
does not contains tedious programming details.

The global scheme used to construct scanners in the Computer Language
Factory is inherited from Metal. We have chosen to use an efficient
existing scanner generator (Flex) and to design a protocol to connect
the generated scanners to the rest of the factory. This view is
different from what is done with SDF \cite{k93} or Syn \cite{Syn}
which incorporate specific sections to specify the lexical elements
into the complete syntax definition.

This choice has been made not only because of the efficiency of Lex
generators but also because it is  very flexible. 

Some other important reasons lead our choice. First, we would like
that the generated scanners can be used in different manner, connect
to Centaur (written in Lisp), directly to Prolog or any other system.
Second, we would like to be able to run the scanner alone for
debugging. This important feature was missing in the Metal
implementation in which it is impossible to test a scanner
independently from a parser. It is also missing when Lex is used in
its traditional way, connected to Yacc or Bison. In our appraoch the
scanner and the parser are not linked together and run as different
process. A special mode allows an easy debugging of the scanner.

Unless it is done for Metal, we do not try to generate a skeleton that
must be then modified by a Sed script. This part of building a scanner
with the help of Metal was too tedious and generated also a
lot of problems because it asked to write Sed scripts and not directly
Lex specifications. The generation of a skeleton of this Sed scripts
by Metal is also useless when one wants to modify an initial
specification. Rather than mechanizing half of the job, we
preferred as a first sketch to let the entire responsibility of
specifying the scanner to the user.  We allow the use of plain Lex
specifications, modulo the use of a specialized library that implements
the interface.  The fix part of the proposed Flex skeleton is rather
small, and due to the technology that is used by the parser, there is
no need for special keywords corresponding to entry points of the
parser as it is the case with Metal. So the user has really to specify
only the atoms of its language, this have to be done anyway, and the
keywords of its language.  However, one can imagine to write a
processor that will check that all tokens and keywords used in a
concrete syntax definition is defined in the correspondent Lex
definition.

There is no encoding of keywords as it was done in Metal. The raw
strings of characters are used. Even if this will slow down a little
the communication between the generated scanners with the rest of the
system, this simplify a lot the process of building scanners and
parsers, and in fact we think that data compression is the job of the
protocol that is used, not the job of the user.

Concerning the parsing phase, the extensions we propose have several
goals.  The first ones deal with efficiency and expressiveness. The
last ones deal with a real lack in traditional generated parsers when
there are used in complex environment: the link between the source
text and the rest of the system is in general done in an ad-hoc way.
Here are the goals we want to reach:

\begin{itemize}
\item allow right recursion in grammar rules: right recursion may be
natural when giving a grammar for a particular language, however it may
lead to infinite loops during Prolog execution; by transforming right
recursive rules into non recusive ones we improve the expressiveness
of the formalism,

\item make use of Prolog compiler optimisations: most of modern
Prolog compiler or interpreteur use indexing on some arguments to
improve clause selection. By transforming rules with similar beginning
of their right part, we can take benefit of indexing for better
performance,

\item generate some code to retrive the location of parsing errors in
the text that is analysed,

\item generate some code to allow an easy correspondance between the
constructed prolog term and the source code that is analysed. This
correspondance will be used in subsequent processors, for exemple
compilers or translators to produce accurate error messages,

\item provide a way to easily keep or discard comments.

\end{itemize}

Our extension of DCGs uses in fact the Prolog syntax and is translated
by a preprocessor to pure DCGs. There is no static analysis as there is
no abstract syntax declaration.

The current version version of the CLF has been used to generate a
parser for XML, then to extend this parser to produce a rule language
that manipulate XML expressions with logical variables. Its appears
that it was simpler and safer to generate a new parser for XML, and then
to modify this parser than to use an existing one.

\htmladdimg{../pics/RBline.gif}

\section{Tutorial}

\htmladdimg{../pics/RBline.gif}

\subsection{Defining a scanner}

\htmladdimg{../pics/GBline.gif}

Defining a scanner for the CLF is not very different from defining a scanner
with Flex only. Only two constraints must be followed:

\begin{enumerate}

\item A skeleton that includes the interface must be used.

\item A library must be used to ``emit'' the tokens that are identified.

\end{enumerate}

The output of the scanner will distinguish the different
\index{token}tokens by their kinds.  There are three basic kinds of
tokens:

\begin{description}

\item[keyword] \index{keyword}This kind contains the reserved words of the language that is 
analyzed and the punctuation.

\item[itoken] \index{itoken}Integer tokens are all the tokens whose values are integers. 

\item[stoken] \index{stoken}String tokens are all the tokens whose values are strings, such
as identifiers for example.

\end{description}

They may be several classes of integer or string tokens. We will see that they
will be distinguished by a \index{tokens name}name.

Three more kinds of tokens are used for practical reasons:

\begin{description}

\item[nl] To signal \index{nl}\index{newline}new lines, so it will be 
possible to count them and report a line number when needed.

\item[comtext] \index{comtext}To identify the text of a \index{comment}comment,
when comments must be kept fir use with a syntactical editors for
example.

\item[error] \index{error}To report an erroneous token that cannot be 
identified by the scanner.

\end{description}

The following sections explain how to construct the lex specification
to build a scanner. They also give some basic and useful examples. More
information on how to build a Flex scanner must be found in the Flex
manual.

\htmladdimg{../pics/GBline.gif}

\subsubsection{The Basic Skeleton\label{The Basic Skeleton}}

\htmladdimg{../pics/GBline.gif}

All Flex specifications used to generate a scanner for the Computer
Language Factory must follow the following \index{basic skeleton}skeleton.

\index{Tokens-flex-define.c} 
\index{Tokens-flex-proto.c}
\begin{quote}
\begin{verbatim}
%{
#include "Tokens-flex-define.c"
%}
%S waitdata sentences comment comm incomm
   <regexp abbreviations>
%%
#include Tokens-flex-fixpart
   <comments definition>
   <tokens definition>
%%
#include "Tokens-flex-proto.c"
   <C user's functions>
\end{verbatim}
\end{quote}

This skeleton already imports ({\tt \#include}) two files
(Tokens-flex-define.c and Tokens-flex-proto.c) used by the Tokens
environment. This environment defines some functions that must be used
to return the tokens that has been identified.


The interface with the host system uses a fix part that must replace
the line {\tt \#include Tokens-flex-fixpart}. As includes are not
allowed in the body of a Flex definition, the job can be done by hand
or with the help of a preprocessor.


The different sections are described in detail below.

\htmladdimg{../pics/GBline.gif}

\noindent {\bf \htmladdimg{../pics/pointa.gif}
Contexts\label{context}}

The following contexts must be present by default.

\index{context}
\index{waitdata}
\index{sentences}
\index{comment}
\index{comm}
\index{incomm}
\begin{quote}
\begin{verbatim}
%S waitdata sentences comment comm incomm
\end{verbatim}
\end{quote}

If some more contexts are necessary to define the scanner of an object
language, some new contexts can be added to this list.

The context ``waitdata'' \index{waitdata}is only used by the
interface. It is use to indicate some options and modes to the scanner
that is generated. The users should not use this context.

The context ``sentences'' \index{sentences}is the context in which
phrases of the object language must be scanned. Most of the job of the
scanner will be done here.

The context ``comment'' \index{comment}is the context in which
comments that are found in an object program text must be
scanned. When the generated scanner identifies the beginning of a
comment, it must switch to this context.

The contexts ``comm'' and ``incomm'' \index{comm}\index{incomm}are
used to analyze comments that are not embedded in a complete program
(i.e. when one want to analyze only the text of a comment). These two
contexts was only used by Centaur to edit comments.

Examples for the two mostly used forms of comments are given later on.

\pagebreak


\htmladdimg{../pics/GBline.gif}

\noindent {\bf \htmladdimg{../pics/pointa.gif}
Regular Expression Abbreviations\label{Regular Expression Abbreviations}}

As in every Flex specification the user can define some abbreviations
that will be used in the rest of the specification. One will refer to
the Flex manual for more information.

The proposed treatments for C-like comments will define some regular
expression abbreviations.


\htmladdimg{../pics/GBline.gif}

\noindent {\bf \htmladdimg{../pics/pointa.gif}
Fix Tokens Interface\label{Fix Tokens Interface}}

The same scanner can be used in different ways: connected to the
Centaur system, connected directly to Prolog to execute a Typol
specification, or in a debug mode. It can also be used to analyze a
complete file or only a given piece of text.

To achieve this, the context ``waitdata'' \index{waitdata}will
recognize some keywords used by the host system to identify itself. It
can also set some variables that will be used by the syntactic
analyzer.

The line {\tt \#include Tokens-flex-fixpart} must be replaced by some
fix Flex definitions.  A {\tt ed } script to perform this
transformation can be found in the Tokens lib directory and is called
by the suggested Buildfile. Be aware that due to this inclusion, error
messages produced by Flex will be shifted (by 20 lines when done by
the ed script).

As the generated scanner may be used in a server mode, the end-of-file
character is not sufficient to mark the end of inputs. A cryptic
keyword that we hope will never appear in programs is used (we can
suspect some bootstapping problems here!). See details in the ``Stans
Alone Exection'' section.


\htmladdimg{../pics/GBline.gif}

\noindent {\bf \htmladdimg{../pics/pointa.gif}
Comments Definition\label{Comments Definition}}

This \index{comment}part may be omitted in a first attempt to define a lexical
analyzer. However to analyse real programs it should be defined anyway.

Two very usual classes of comments are defined somewhere else in this
document as examples. One can adapt them very easily.

These two examples are:

\begin{description}

\item[C-like Comments] \index{C-like comments}These comments can be found 
every where in the text.
They start with some particular string (``{\tt /*}'' in C) and end with an
other one (``{\tt */}'' in C). They may spread over several lines.


\item[ADA-like Comments] \index{ADA-like comments}These comments start 
with some particular string 
(two ``{\tt -}'' in ADA) somewhere on the line and end at the end of the line. 
They cannot spread over several lines.

In the current version of Tokens, comments are passed over to the
Centaur system, while it is not the case when the scanner is connected
directly to Prolog.

\end{description}

\htmladdimg{../pics/GBline.gif}

\noindent {\bf \htmladdimg{../pics/pointa.gif}
Tokens Definition\label{Tokens Definition}}

This is the real part of the scanner definition. It must contain the
definition of the \index{token}\index{keyword}keywords (including
punctuation and other signs) and the definition of all the tokens of
the language (identifiers, numbers, strings, etc.). Tokens come in
various classes, depending on the concrete syntax definition, that are
identified by their \index{tokens name}names, for example {\tt ID} for
identifiers or {\tt UCID} for identifiers starting with an uppercase
letter.

When a token as been identified, one of the function provided by the
library must be used to return its value (the ``return'' action
traditionally used in Flex must not be used directly):

\begin{description}

\item[keyword(char* key)] \index{keyword}The string passed as argument 
must represent
the keyword itself, no encoding is used as it was the case in Mentor
or Centaur. The string must be the same that those that will be used
in the parser definition.

\item[itoken(char* tok, char *value), stoken(char* tok, char *value)] 
\index{itoken}\index{stoken}
The first function 
must be used to return tokens whose value is an integer, the second to return 
tokens whose value is a string. 

The string {\tt tok} is a \index{tokens name}token name. The same name
must be used in the parser definition.

The string {\tt value} is the value of the token. It is often useful
to put this value in a canonical form, suppressing escape characters
and boundaries. For example, the value of a string written ``{\tt
"aaa""bbb"}'' in the text of a program may be simply ``{\tt
aaa"bbb}''. Keeping this form in the abstract syntax tree will make
easier the writing of the semantics of the language, in particular
when one wants to write translations. Of course, the un-parser will
have to take this into account, adding necessary quoting and escape
characters. In the case of a non case-sensitive language, it will be
also useful to normalize the identifiers, putting them in lowercase
for example. Doing this, the semantics will not have to manipulate
strings when comparing the equality of to strings.

\end{description}


\htmladdimg{../pics/GBline.gif}

\noindent {\bf \htmladdimg{../pics/pointa.gif}
New Lines\label{New Lines}}

Every new lines must be reported. This will allow reporting of errors
during parsing. 

This information is also used to make a correspondence between an
occurence in the abstract syntax tree and the line number when one
wants to execute and off-line version of the semantics.

At each new line, a call to the function \index{nl()}{\tt nl()} must
be done. Be careful to report also new lines that occur inside some
tokens (in particular comments), otherwize line numbers reported by the
parser or by the semantics will be erroneous.


\htmladdimg{../pics/GBline.gif}

\noindent {\bf \htmladdimg{../pics/pointa.gif}
C User's Functions\label{C Users's Functions}}

\index{user's functions}
As in every Flex definition, this part may contain C functions that
are need by the scanner. They may be functions used to put some tokens
in a canonical form, suppressing escape characters or boundaries for
example. A catalogue of such functions for languages already used in
Centaur is provided in the library.


\htmladdimg{../pics/GBline.gif}

\subsubsection{Example}

\htmladdimg{../pics/GBline.gif}

We will take as example the traditional Exp language, taken from the
Centaur documentation.

This formalism uses some keywords (arithmetic signs), identifiers
(let's say that they must start with uppercase letters, then continue
with uppercase letters, digits or underscores) and integers. These
tokens will be classified into 2 classes named {\tt ID} and {\tt INT}.

The two regular expressions that recognize our tokens are:

\begin{itemize}

\item {\tt [A-Z][A-Z0-9\_]*}

\item {\tt [0-9][0-9]*}

\end{itemize}

Identifiers will be represented as strings in the abstract tree, while
integers will be represented as integers. Thus there will be reported
using respectively the functions {\tt stoken} and {\tt itoken}.

Comments are ADA-like comments starting with the {\tt \%} sign. They
end at the end of the line

Placing these elements in the skeleton give us the following
definition:

\begin{quote}
\begin{verbatim}
%{
#include "Tokens-flex-define.c"
%}
%S waitdata sentences comment comm incomm
%%
#include Tokens-flex-fixpart

<comment>.*		comtext(yytext);
<comment>\n[ ]*\n	{BEGIN sentences;nl();nl();}
<comm>"%"\n 	{comtext("");};
<comm>"%" 	{BEGIN incomm; };
<incomm>.* 	comtext(yytext);
<comm,incomm>\n {BEGIN comm; nl(); }
<comm,incomm>[ \t] ;

<sentences,comment>\n	{BEGIN sentences; nl(); }
<sentences,comment>[ \t]	;

<sentences>"%"\n  	{comtext("");nl();};
<sentences>"%"		{BEGIN comment;};

"=" |
"+" |
"-" |
"*" |
"(" |
")" |
";"             keyword(yytext);


<sentences>[A-Z][A-Z0-9_]*      stoken("ID",yytext);
<sentences>[0-9][0-9]*          itoken("INT",yytext);
%%
#include "Tokens-flex-proto.c"

\end{verbatim}
\end{quote}

\htmladdimg{../pics/GBline.gif}

\subsubsection{C-like Comments\label{C-like Comments}}

\htmladdimg{../pics/GBline.gif}

\index{C-like comments}
C-like comments starts with the two characters ``/*'' and end with
``*/''. The following definition considers each line as one
token. New-lines are not considered as part of the token. In this
version, leading spaces at the beginning of the commens are removed.

\begin{quote}
\begin{verbatim}
C1	[^*\n]
C2	[^*/\n]
Ca	({C1}*("*"*{C2})?)*
Cb	({C1}*("*"*{C2})?)*"*"*
\end{verbatim}
\end{quote}

\begin{quote}
\begin{verbatim}
<comment>{Ca}\n	 |
<comment>{Cb}\n  {suplast(); rmtabs(); comtext(yytext); nl();};
<comment>{Cb}"/" {BEGIN sentences; suplast2(); rmtabs(); \
                  comtext(yytext);};

<sentences>"/*"  {BEGIN comment; };

<comm>.*         {rm_leading_spaces(); comtext(yytext);};
<comm>\n         nl();
\end{verbatim}
\end{quote}

\htmladdimg{../pics/GBline.gif}

\subsubsection{ADA-like Comments\label{ADA-like Comments}}

\htmladdimg{../pics/GBline.gif}

\index{ADA-like comments}
ADA-like comments start with two minus signs anywhere on the line and
stop at the end of the line. In the following definition, the value of
a comment does not include the two minus signs, neither the last
new-line character.

\begin{quote}
\begin{verbatim}
<comment>.*        comtext(yytext);
<comment>\n[ ]*\n  {BEGIN sentences; nl(); nl();}
<comm>--\n         {comtext("");nl()};
<comm>"--"         {BEGIN incomm; };
<incomm>.*         comtext(yytext);
<comm,incomm>\n    {BEGIN comm; nl(); }
<comm,incomm>[ \t] ;

<comment>\n	   {BEGIN sentences; nl(); }
<comment>[ \t]	   ;

<sentences>--\n    {comtext(""); nl();};
<sentences>"--"    {BEGIN comment;};
\end{verbatim}
\end{quote}

\htmladdimg{../pics/GBline.gif}

\subsection{Defining a parser}

\htmladdimg{../pics/GBline.gif}

Defining a parser in CS is very similar to using traditional DCGs.
However, the sign {\tt -->} is replaced by {\tt :-} as in regular
prolog clauses.  Every DCGs documentation can thus be used.

\htmladdimg{../pics/GBline.gif}

\subsubsection{Grammar rules}

\htmladdimg{../pics/GBline.gif}

The following exemple shows an exemple of rules:

\begin{quote}
\begin{verbatim}
exp(plus(A,B)) :- exp(A), ['+'], exp(B).
exp(minus(A,B)) :- exp(A), ['-'], exp(B).
\end{verbatim}
\end{quote}

As explain in the rational, right recursion is permited, as it is the
case in this example. Moreover, the generated code will take advantage
of the regularity of these rules to generate indexable clauses if necessary.

Unless traditional DCGs, all predicates must have an argument that
explain how to construct the resulting term.

Notice that some production rules may only pass a term, not
constructing a new one. This is in particular the case when production
rules are used to manage prorities of syntactic operators for example:

\begin{quote}
\begin{verbatim}
exp(A) :- factor(A).
\end{verbatim}
\end{quote}

Lists are treated as ordinary prolog lists.

\begin{quote}
\begin{verbatim}
element_s([E|L]) :-
	element(E), element_s(L).
element_s([]).
\end{verbatim}
\end{quote}

\htmladdimg{../pics/GBline.gif}

\subsubsection{Tokens}

\htmladdimg{../pics/GBline.gif}

The special literal {\tt stoken} must be used to connect the parser to
the scanner. Its first argument must the name of the token as it is
given in the scanner. The second argument is of the form {\tt
string(A)} or {\tt integer(A)} depending on the type of the token.

Here is some examples.
\begin{quote}
\begin{verbatim}
exp(id(A)) :- stoken('IDENTIFIER',string(A)).
exp(text(A)) :- stoken('TEXT',string(A)).
exp(int(A)) :- stoken('INTEGER',integer(A)).
\end{verbatim}
\end{quote}

Theses rules correspond respectively to the following ones in the Lex
definition:

\begin{quote}
\begin{verbatim}
stoken("IDENTIFIER",yytext)
stoken("TEXT",yytext);
itoken("INTEGER",yytext);
\end{verbatim}
\end{quote}

\htmladdimg{../pics/GBline.gif}

\section{User's Manual}

\htmladdimg{../pics/RBline.gif}

\subsection{Producing a scanner}

\htmladdimg{../pics/GBline.gif}

When a Flex specification for the CLF has been written, it must be
compiled as any other Flex specification. The C code that is produced
will be then compiled with a C compiler.

The scanner that is generated can be used independently for
debugging, or used with a direct connection with Prolog.

\htmladdimg{../pics/GBline.gif}

\subsubsection{Files Naming Convention}

\htmladdimg{../pics/GBline.gif}

\index{naming convention}

Suppose that we want to define a scanner for a formalism named {\tt
Foo} and that this scanner will be a component of a concrete syntax
definition of {\tt Foo} named {\tt std} (remember that they may be
several concrete syntax for a unique abstract syntax). The Flex
definition of the scanner must be found in the file {\tt
Foo-std-Tokens.lex} and the generated scanner will be found in {\tt
Foo-std-Tokens}.

\htmladdimg{../pics/GBline.gif}

\subsubsection{Compiling Flex Specifications\label{Compiling Flex Specifications}}

\htmladdimg{../pics/GBline.gif}

\index{compiler}
As usual, Flex specifications must be compiled by Flex, and the ouput
produced by Flex compiled again by a C compiler.

The following example of Makefile \index{Makefile} will compile two
scanners (named {\tt std} and {\tt spc}) for the language {\tt Foo}.

\begin{quote}
\begin{verbatim}
all: Foo-std-Tokens Foo-spc-Tokens

include ..../clf//Tokens/BuildTokens
\end{verbatim}
\end{quote}

The path .../clf/Tokens/lib must be set accordingly with the
installation directory of CLF.

Be aware that due to the use of a preprocessor to include the fix part
of the interface, error messages produced by Flex will be shifted (by
20 lines).

\htmladdimg{../pics/GBline.gif}

\subsubsection{Stand Alone Execution}
\index{Stand Alone Execution}

\htmladdimg{../pics/GBline.gif}

The generated scanners can be run in a stand alone 
\index{debug mode}mode for testing and debugging. Following our example, one can
run the generated scanner typing {\tt ./Foo-std-Tokens}, then the
keyword
\index{BEGINDATA}{\tt [BEGINDATA]} alone on one line. The scanner is
ready to scan what arrives on its input and produces on its output one
line for each token that is recognized, containing all information
about the token (type, name, value). It will stop scanning when the
string {\tt [)*\&\^\ \%!ENDDATA!\%\&*(]} is entered alone on a line. It
will restart scanning by typing {\tt [BEGINDATA]} again.

It is also possible to scan a complete file by entering
\index{PARSEFILE}{\tt [PARSEFILE]} immediately followed by the name of
a file on a single line.

The exact protocol that is sent to Centaur or Prolog can be viewed by
typing \index{TARGET}\index{Centaur mode}{\tt [TARGET]centaur} or
\index{Eclipse mode}{\tt [TARGET]eclipse} alone on one line before
entering the command {\tt [BEGINDATA]}.

The scanner can be terminated typing the command \index{QUIT}{\tt
[QUIT]}. It is also possible to abort a scanner as every process.

\htmladdimg{../pics/GBline.gif}

\subsection{Producing a parser}

\htmladdimg{../pics/GBline.gif}

When a CS specification has been written, it must be compiled to
Prolog DCGs.

The generated code can be used directly by a prolog program.

\htmladdimg{../pics/GBline.gif}

\subsubsection{Files Naming Convention}

\htmladdimg{../pics/GBline.gif}

Suppose that we want to define a scanner for a formalism named {\tt
Foo}, and that we want to define the standard parser for this
formalism. The CS definition of the parser must be found 
in a file named {\tt Foo-std.csp}. When compiled to DCGs a Prolog file named
{\tt Foo-std.sp} will be created.

\htmladdimg{../pics/GBline.gif}

\subsubsection{Compiling a Specification}

\htmladdimg{../pics/GBline.gif}

The following example of Makefile \index{Makefile} will compile
a parser define in {\tt Foo-std.csp}.

\begin{quote}
\begin{verbatim}
all: Foo-std.sp

expand_cs = .../clf/Syntax/expand_cs

Foo-std.sp : Foo-std.csp
	${expand_cs} Foo-std.csp Foo-std.sp 


\end{verbatim}
\end{quote}

The path .../clf/Tokens/lib must be set accordingly with the
installation directory of CLF.

\htmladdimg{../pics/GBline.gif}

\subsubsection{Using a parser}

\htmladdimg{../pics/GBline.gif}

To use a parser, one must define the following predicate in which
``Foo'' is the name of the formalism, ``std'' the name of the parser,
and entry the name of the top production rule to be used.

\begin{quote}
\begin{verbatim}
'Foo-std-parser'(_A, __A) -->
   skip_comm(0, _1),
   entry(_A, __A, _1, _2),
   skip_comm(_2, _).
\end{verbatim}
\end{quote}

The call to the parser itself will be defined as follow. The input of
the predicate is the name of the file to be parsed. They are two
results, a tree which is the abstract tree, and a special one used to
make a correspondnance between occurences in the parsed tree and line
numbers in the source.

\begin{quote}
\begin{verbatim}
readfile(File,Tree,Refs) :-
  clf_parse(``/home/users/../Foo-std-Tokens'',
            'Foo-std-parser',
            File,
            Tree, Refs).
\end{verbatim}
\end{quote}

The first argument of the predicate {\tt clf\_parse} is an absolute
path to the lexical analysor produced by Flex. The second one is the
name of the predicate to be caller, defined before.

Several files must be loaded (compiled) into prolog to get a complete parser.

\begin{quote}
\begin{verbatim}
:-compile("/home/.../clf/Parser/lex_caller.sp").
:-compile("/home/.../clf/Parser/clf_parser.sp").
:-compile("/home/.../clf/Parser/skip_comm.sp").


:-compile("/home/.../Foo/syntax/Foo-std.sp").
:-compile("/home/.../Foo/syntax/Foo-std-parser.sp").
\end{verbatim}
\end{quote}

The different paths must be set accordingly to the installation.

\htmladdimg{../pics/GBline.gif}

\section{Reference Manual}

\htmladdimg{../pics/RBline.gif}

The generation of lexical analyzers relies on a C library used by Flex
that are detailed in the two following sections. They will be founded
in the directory .../Tokens/lib. 

\htmladdimg{../pics/RBline.gif}

\subsection{Tokens Basic Library\label{Tokens Basic Library}}

\htmladdimg{../pics/GBline.gif}

The file {\tt Tokens-flex-proto.c} contains the functions that must be
used to return tokens from the generated scanner.

Note that when Flex recognizes a token, its text value is kept in
a string variable named {\tt yytext}.

\begin{description}

\item[nl()] Indicates that a new line must be taken into account.

\item[comtext(char *a)] Report the value of a comment. Don't forget that if
you decide that boundaries are not part of the value, they must be added by the
un-parser.

\item[stoken(char* tok, char *value)] Report the value of a token implemented 
as string. The first argument is a string containing the class name of the
token, the second one its value. To make easier the writing of semantics tools
it is recommended that the values of tokens do not contain boundaries, neither
escape character. In this case these must be properly added by the un-parser
to generate back a correct text.

\item[itoken(char* tok, char *value)] Report the value of a token implemented 
as integer. The first argument is a string containing the class name of the
token, the second one its value. 

\item[keyword(char *key)] Report a token of class keyword. The argument is
a string representing the keyword itself.

\item[error(char *val)] Report an erroneous token.

\end{description}

\htmladdimg{../pics/GBline.gif}

\subsection{Functions on Strings\label{Functions on Strings}}

\htmladdimg{../pics/GBline.gif}

The file {\tt Tokens-flex-proto.c} \index{Tokens-flex-proto.c}also
defines some functions that can be used to manipulate strings before
returning them. They modify directly the value of the variable {\tt
yytext} \index{yytex} that Flex uses to keep the value of the tokens
that have been recognized.

\begin{description}

\item[shiftleft()] \index{shiftleft()}Suppress the first character from {\tt yytext}.

\item[shiftleft2()] \index{shiftleft2()}Suppress the two first characters from {\tt yytext}.

\item[suplast()] \index{suplast()}Suppress the last character from {\tt yytext}.

\item[suplast2()] \index{suplast2()}Suppress the two last characters from {\tt yytext}.

\item[rmtabs()] \index{rmtabs()}Replace tabs by spaces in {\tt yytext}.

\item[rm\_leading\_spaces()] \index{rm\_leading\_spaces()}Suppress spaces at the beginning of {\tt yytext}.

\item[clear\_string(char c)] \index{clear\_string(char c)}Suppress any forcing character c from {\tt yytext}.

\item[clear\_string1()] \index{clear\_string1()}Define the first character as beeing a forcing character, then suppress the first and the last characters and also any occurence of the forcing character from {\tt yytext}. For example tranform the string {\tt 'foo''bar'} into {\tt foo'bar}.

Example of use to define prolog-like atoms:

\begin{verbatim}
<sentences>'([^'\n]*|'')*'   {clear_string1();stoken("ATOM",yytext);};
\end{verbatim}

\item[clear\_string2()] \index{clear\_string2()}Suppress the first and the last characters and also any occurance of the forcing character backslash in {\tt yytext}.

\end{description}

\htmladdimg{../pics/GBline.gif}

\subsection{Retrieving line numbers in the source}

\htmladdimg{../pics/GBline.gif}

The last argument of the predicate {\tt clf\_parse} returns a term that
permit to make the correspondance between the parsed term and the
source code. The structure reflects exactly the structure of the
parsed term, but there is only one operator ({\tt node}) that has one more son
that contains a line number.

For example, if the source text, stating at line 12 is

\begin{quote}
\begin{verbatim}
A + 
  B
\end{verbatim}
\end{quote}

the construted term will be

\begin{quote}
\begin{verbatim}
plus(id('A'),id('B'))
\end{verbatim}
\end{quote}

and the reference term will be

\begin{quote}
\begin{verbatim}
node(12,node(12,nil),node(13,nil))
\end{verbatim}
\end{quote}

\htmladdimg{../pics/GBline.gif}

\section{Implementation Notes}

\htmladdimg{../pics/RBline.gif}

The goal of this section is to give an idea of the transformations that
are performed on CS rules to produce standard DCG rules.

We focus on two transformations that are used to take advantage of
clauses indexing and to allow left recursion. Of courses these two
transformationsshould be combined in some cases. We also focus only on
the basic transformations, not taking into account additionnal
parameters that are generated for other reasons.

When rules can be chosen by use of a discriminent token as in

\begin{quote}
\begin{verbatim}
a(X) :- ['t1'], b(X).
a(X) :- ['t2'], b(X).
...
\end{verbatim}
\end{quote}

the following code is produced

\begin{quote}
\begin{verbatim}
a(X) -> [T], a$h(T,X).
a$h('t1',X) -> b(X).
...
\end{verbatim}
\end{quote}

We have chosen to perform the transformation only when 3 different
tokens are used.

When the following schema is used

\begin{quote}
\begin{verbatim}
add(plus(A,B)) :- add(A), ['+'], mult(B).
add(X) :- mult(X).
\end{verbatim}
\end{quote}

the following code is produced

\begin{quote}
\begin{verbatim}
add(C) -> mult(A), add$x(A,C).
add$x(A,C) -> ['+'], add$x(plus(A,B),C).
add$x(A,A) -> .
\end{verbatim}
\end{quote}

CS definition uses one parameter (the constructed term). The generated
DCGs rules contain 4 ou 5 parameters. When extended again to prolog
clauses this gives 6 or 7 parameters for each clause.

\bibliography{clf_manual}
\bibliographystyle{abbrv}


\end{document}